\documentclass[epj]{svjour}
\begin{document}
\title{State-dependent  graviton noise in the equation of geodesic deviation}

\author{Z. Haba}
\offprints{Z. Haba}          
\institute{Institute of Theoretical Physics, University of Wroclaw, 50-204 Wroclaw,email:zbigniew.haba@uwr.edu.pl}

\date{Received: date / Revised version: date}
%
\abstract{We consider an  equation of the geodesic deviation
appearing in the problem of gravitational wave detection in an
environment of gravitons. We investigate a state-dependent
graviton noise ( as discussed in a recent paper by Parikh,Wilczek
 and Zahariade) from the
point of view of the Feynman integral and
 stochastic differential equations. The evolution of the density
 matrix  and the transition probability in an environment of gravitons is obtained.  We express
 the time evolution by a solution of a stochastic geodesic deviation equation with
 a noise dependent on the quantum state of the gravitational
 field.}

\PACS{ {PACS-key}{} \and
             {PACS-key}{}}
\maketitle

\section{Introduction}
The study of the effect of an environment, which is not directly
observable, on the motion of macroscopic bodies, began with the
phenomenon of Brownian motion. The mathematical theory of Brownian
motion allowed to confirm the presence of invisible molecules (
see \cite{nelson}). In a recent paper Parikh,Wilczek
 and Zahariade (PWZ) \cite{wilczek} suggest that in a similar way we can detect the
 environment of gravitons in spite of the negative conclusions of
 Dyson \cite{dyson}. According to PWZ the noise resulting from
 the ground state or coherent state of gravitons is weak and
 practically undetectable. However, the high temperature
states  and squeezed states of the graviton can lead to a
substantial increase of  noise which can be detected in the
gravitational wave detection experiments. The authors
\cite{wilczek} derive their results by an investigation of the
geodesic deviation equation in the environment of the quantized
gravitational field. They apply the influence functional method
\cite{feynman} in order to transform the evolution of the
transition probability of macroscopic bodies into an expectation
value with respect to the noise disturbing the geodesic deviation
equation. Studies on the effect of gravitons on the motion of
other particles appeared earlier
\cite{hu}\cite{habampl}\cite{hk}\cite{hu2}\cite{jackel}\cite{prd},
but these papers did not tackle directly the geodesic deviation
equation. The quantum noise from the environment has  been
investigated in other fields of physics \cite{kac}\cite{leggett}.
It can be detected in quantum optics (ref. \cite{lax}, sec.14) and
solid state physics.

The evolution of the density matrix in an environment of
unobservable particles is usually treated by means of the
influence functional \cite{feynman} which starts from the Feynman
integral. As a result one can express the evolution of the density
matrix as a solution of the master equation. One can also express
the density matrix  by the Wigner function and derive a stochastic
equation for the evolution of the Wigner function. We have derived
such equations for the thermal graviton environment by means of
the Feynman integral in \cite{hk}. In this paper we apply the
method to the quantum evolution of a particle following the
geodesic deviation equation when the quantum gravitational field
is in an arbitrary Gaussian  state. We apply a method of
transforming the Feynman integral into an expectation value with
respect to the Brownian motion developed in
\cite{habajpa}\cite{hababook}. In such a case the dependence of
the evolution of the density matrix on the state of the
gravitational field is exhibited explicitly.

The plan of the paper is the following. In sec.2 we derive in a
novel way the transformation of the Feynman integral in quantum
mechanics into an expectation value over a state-dependent noise.
In sec.3 we show how this transformation applies to quantum field
theory. In sec.4 we discuss in detail perturbations by noise
resulting from  the time-dependent Gaussian states. We explain the
 calculations  of the density matrix and the
transition probability in a linear coupling to the environment in
sec.5. We apply the method to the PWZ model in sec.6 (as
introduced in \cite{wilczek}) in the one mode approximation. In
sec.7 the influence of infinite number of modes of the thermal
gravitational field upon  the geodesic deviation equation is
treated by the method of the forward-backward Feynman integral as
previously applied in \cite{hk} to a geodesic motion. In sec.8 we
apply our method of the state-dependent noise to derive a
stochastic deviation equation which governs the evolution of the
density matrix and the transition probability. Sec.9 contains a
summary of the results. In the Appendix we discuss the stochastic
deviation equation which results from the assumption that the
initial values of the classical gravitational field have the
thermal Gibbs distribution.
\section{State-dependent transformation of the Feynman integral}
Let us consider first a simple model of the Schr\"odinger equation
of quantum mechanics in one dimension with the time-dependent
potential $U_{t}$
\begin{equation}
i\hbar\partial_{t}\psi_{t}=(-\frac{\hbar^{2}}{2m}\nabla_{x}^{2}+U_{t})\psi_{t}.
\end{equation}
Assume that we have a solution $\psi_{t}^{g}$ of another
Schr\"odinger equation (with a potential $\tilde{U}_{t}$)
\begin{equation}
i\hbar\partial_{t}\psi_{t}^{g}=(-\frac{\hbar^{2}}{2m}\nabla_{x}^{2}+\tilde{U}_{t})\psi_{t}^{g}.
\end{equation}
Let us write the solution of eq.(1) in the form
\begin{equation}
\psi_{t}=\psi_{t}^{g}\chi_{t}.
\end{equation}
Inserting $\chi_{t}$ from eq.(3) into eqs.(1)-(2) we find that
$\chi_{t}$ satisfies the equation
\begin{equation}\begin{array}{l}
\partial_{t}\chi_{t}=\frac{i\hbar}{2m}\nabla_{x}^{2}\chi_{t}+\frac{i\hbar}{m}
(\nabla_{x}\ln\psi_{t}^{g})\nabla_{x}\chi_{t}-\frac{i}{\hbar}(U_{t}-\tilde{U}_{t})\chi_{t}
\end{array}\end{equation}
with the initial condition
\begin{equation}
\chi_{0}=\psi_{0}(\psi^{g}_{0})^{-1}
\end{equation}
expressed by the initial conditions for $\psi_{t}$ and
$\psi_{t}^{g}$.

We apply eq.(4) either with $\tilde{U}_{t}=U_{t}$ or with $
U_{t}=V_{t}+\frac{m\omega^{2}x^{2}}{2}$ and
$\tilde{U}_{t}=\frac{m\omega^{2}x^{2}}{2}$. In the first case
eq.(4) reads
\begin{equation}\begin{array}{l}
\partial_{t}\chi_{t}=\frac{i\hbar}{2m}\nabla_{x}^{2}\chi_{t}+\frac{i\hbar}{m}
(\nabla_{x}\ln\psi_{t}^{g})\nabla_{x}\chi_{t}.
\end{array}\end{equation}
In the second case
\begin{equation}\begin{array}{l}
\partial_{t}\chi_{t}=\frac{i\hbar}{2m}\nabla_{x}^{2}\chi_{t}+\frac{i\hbar}{m}
(\nabla_{x}\ln\psi_{t}^{g})\nabla_{x}\chi_{t}-\frac{i}{\hbar}V_{t}\chi_{t}
\end{array}\end{equation}
Eq.(6) can be considered as the diffusion equation with the
imaginary diffusion constant $\frac{i\hbar}{m}$ and a
time-dependent drift $\frac{i\hbar}{m} \nabla_{x}\ln\psi_{t}^{g}$.
If we extend the diffusion theory \cite{nelson}\cite{freidlin} to
a complex domain then we can conclude that the solution of eq.(6)
is determined by the solution of the Langevin equation
\begin{equation}
dq_{s}=\frac{i\hbar}{m}\nabla\ln\psi_{t-s}^{g}(q_{s})ds+\sqrt{\frac{i\hbar}{m}}db_{s}.
\end{equation}
Here, the Brownian motion $b_{s}$ is defined as the Gaussian
process with the covariance
\begin{equation}
E[b_{t}b_{s}]=min(t,s).\end{equation} The solution of eq.(6) is
\begin{equation}
\chi_{t}(x)=E\Big[ \chi_{0}(q_{t}(x))\Big],
\end{equation}
where $q_{t}(x)$ is the solution of eq.(8) with the initial
condition $q_{0}(x)=x$ and the expectation value is over the paths
of the Brownian motion.

The solution of eq.(7) is expressed \cite{freidlin} by the
Feynman-Kac formula (where the Feynman paths are replaced by the
paths of the diffusion process $q_{s}(x)$ of eq.(8))
\begin{equation}
\chi_{t}(x)=E\Big[\exp\Big(-\frac{i}{\hbar}\int_{0}^{t}dsV_{t-s}(q_{s}(x))\Big)\chi_{0}(q_{t}(x))\Big]
\end{equation}

 We could derive eq.(10) from the Feynman integral. The solution $\psi_{t}$ of eq.(1) with the initial
condition $\psi_{0}$ at $t=0$ can be expressed by the Feynman
integral
\begin{equation}\begin{array}{l}
\psi_{t}(x)=\int {\cal
D}q\cr\exp\Big(\frac{i}{\hbar}\int_{0}^{t}\Big(\frac{1}{2}m(\frac{dq_{s}}{ds})^{2}-U_{t-s}(q_{s})
\Big)ds\Big)\psi_{0}(q_{t}(x)),
\end{array}\end{equation} where the integral is over paths $q_{s}(x)$ starting from $x$ (i.e.$q_{0}(x)=x$; we
shall also denote $q_{s}(x)$ by $q_{s}$ when there is no danger of
confusion).

We show (first for the formula (10)) that if in the Feynman
formula (12) we take as Feynman paths the paths of the diffusion
process (8) then we obtain eq.(10). The proof is based on the
representation of
$\psi_{0}^{g}(q_{t})=\exp\ln\Big(\psi_{0}^{g}(q_{t})\Big)$ as an
integral
\begin{equation}
\ln\psi_{0}^{g}(q_{t}(x))=\ln\psi_{t}^{g}(x)+\int_{0}^{t}
d\ln\psi_{t-s}^{g}(q_{s}(x))
\end{equation}
and the identity (for the process (8))
\begin{equation}\begin{array}{l}
\frac{i}{\hbar}\int_{0}^{t}\Big(\frac{1}{2}m(\frac{dq_{s}}{ds})^{2}-U_{t-s}(q_{s})\Big)+\ln(\psi_{0}^{g}(q_{t}(x))
\cr=-\frac{1}{2}\int_{0}^{t}(\frac{db}{ds})^{2}+\ln\psi^{g}_{t}(x).
\end{array}\end{equation}
Eq.(14) shows that the integration over $q$ in the Feynman formula
(12) can  be replaced by an average over Brownian paths $b_{t}$
because we arrive at eq.(10) (which still will be proved in
another way by means of the stochastic calculus).

 It is crucial
for the derivation that functionals of the Brownian motion satisfy
a modified differential formula (Ito formula \cite{ikeda}, for an
elementary version see \cite{simon}) following from the
non-differentiability of the Brownian paths $b_{t}$
\begin{equation}\begin{array}{l}
df_{s}(b)=\nabla f_{s}\circ db+\partial_{s}f_{s}ds\cr=\nabla f_{s}
db+\frac{1}{2}\nabla^{2}f_{s}
<(db)^{2}>+\partial_{s}f_{s}ds\cr=\nabla f_{s}
db+\frac{1}{2}\nabla^{2}f_{s} ds+\partial_{s}f_{s}ds,\end{array}
\end{equation}
where $\circ$ denotes the Stratonovitch differential (the
differential $db$ without circle is called Ito differential
\cite{ikeda}) and on the rhs of eq.(15) we insert $<(db)^{2}>=ds$
(we denote $E[..]$ by $<..>$ as a shorthand). If the Langevin
equation (8) is satisfied and if $f_{s}$ is a function of q then
\begin{displaymath} df_{s}=\nabla f_{s}\circ
dq+\partial_{s}f_{s}ds=\nabla f_{s} dq+\frac{i\hbar}{2
m}\nabla^{2}f_{s} ds+\partial_{s}f_{s} ds.
\end{displaymath}
We insert $f_{s}=\ln\psi_{t-s}^{g}$.Then
\begin{equation}
d\ln\psi^{g}_{t-s}(q_{s})=\nabla\ln\psi_{t-s}^{g}dq_{s}+\frac{i\hbar}{2m}\nabla^{2}\ln\psi^{g}_{t-s}ds
+\partial_{s}\ln\psi_{t-s}^{g} ds.
\end{equation}
We have
\begin{displaymath}
\nabla^{2}\ln\psi_{t-s}^{g}=(\nabla^{2}\psi_{t-s}^{g})(\psi_{t-s}^{g})^{-1}-
(\psi_{t-s}^{g})^{-2}(\nabla\psi_{t-s}^{g})^{2}.
\end{displaymath}
We use the fact that $\psi_{t-s}^{g}$ satisfies the Schr\"odinger
equation (2) (with $\tilde{U}=U$) . Then, in equation (16) we can
express $\nabla^{2}\psi_{t-s}$ by $U_{t-s}$ and
$\partial_{s}\psi_{t-s}^{g}$. We can check after an insertion of
 eqs.(13) and (16)  in eq.(14) that
in the exponential of the Feynman integral there remains solely
$-\frac{1}{2}\int ds(\frac{db}{ds})^{2}$. This means that the
functional integral in the Feynman formula (12) is reduced to an
average over the Brownian motion as in eq.(10).

The formula (10) can be proved directly by differentiation (using
the Ito formula). As $q_{s}$ (8) is a Markov process eq.(10)
defines a semigroup. Then, in order to show that the formula (10)
solves  eq.(6) it is sufficient to calculate the generator of the
semigroup at $t=0$. We have from eq.(10)
\begin{equation}\begin{array}{l}
d\chi_{t}=E[\nabla\chi_{0}dq_{s}+\frac{1}{2}\nabla^{2}\chi_{0}dqdq]
\end{array}\end{equation}
In this formula we insert  $dq_{t}$ from eq.(8). Note that
$E[Fdb_{t}]=0 $(if $F$ depends on time less or equal to $t$) and
$dqdq=\frac{i\hbar}{m}dt$. After the calculation of the
differentials we may let $t\rightarrow 0$ to convince ourselves
that the rhs of $d\chi$ is
$\frac{i\hbar}{2m}\nabla^{2}\chi+\frac{i\hbar}{m}\nabla
\ln\psi^{g}_{t}\nabla\chi$.

If in eq.(2) we used
 $\tilde{U}=\frac{m\omega^{2}x^{2}}{2}$
  in eq.(16) with $U=V+\frac{m\omega^{2}x^{2}}{2}$ then after an
insertion of eq.(16) in eq.(12) the potential $U-\tilde{U}=V$
would remain in the Feynman integral. Hence, we would obtain the
result (11). This equation can be proved directly using the
stochastic calculus (eq.(11) defines a semigroup, hence it is
sufficient to calculate its generator at $t=0$)
\begin{equation}\begin{array}{l}
d\chi_{t}=E[-\frac{i}{\hbar}V_{t}(q_{t})\chi_{0}(q_{t})dt+ \nabla
\chi_{0}dq+\frac{1}{2}\nabla^{2}\chi_{0}dq dq].
\end{array}\end{equation}
When we insert $dq_{t}$ from eq.(8)  then we obtain the rhs of
eq.(7).

 The imaginary time version of our formulas
 belongs to
the standard theory of stochastic differential equations
\cite{freidlin}. The derivation of the Feynman integral has been
discussed earlier in a different form in
\cite{habajpa}\cite{hababook}. The stochastic formulas for the
Feynman integral are also discussed in
\cite{doss1}\cite{alb}\cite{doss2}\cite{mazz}. There are some
mathematical subtleties concerning analytic properties of
functions of the stochastic processes as our formalism is an
analytic continuation of the one well known for the imaginary
time. For a large class of analytic potentials the replacement of
the real paths of Feynman by complex diffusion processes driven by
Brownian paths is completely equivalent (this is a mathematical
theory of the Feynman integral). As follows from the derivation of
the Schr\"odinger equation by means of the stochastic calculus the
formula (11) holds true for arbitrary particular solution
$\psi^{t}_{g}$ and arbitrary initial condition
$\psi_{0}^{g}\chi_{0}$ if these functions have a holomorphic
extension from the real line.

If we take $\psi_{t}^{g}$ in the WKB form
$\exp(\frac{i}{\hbar}S_{t})$ then in the limit $\hbar\rightarrow
0$ of the Schr\"odinger equation the function $S$ satifies the
Hamilton-Jacobi equation. The stochastic equation (8)
\begin{displaymath}dq_{s}=-\frac{1}{m}\nabla S_{t-s}(q_{s})ds+\sqrt{\frac{i\hbar}{m}}db_{s}
\end{displaymath}
 together with eq.(10) can be considered as a Hamilton-Jacobi version of a sum over trajectories.

 The formalism can
be extended to arbitrary number of dimensions and to Lagrangians
of the form
\begin{equation}
{\cal L} =\frac{1}{2}g_{lk}\frac{dy^{l}}{ds}\frac{dy^{k}}{ds}-U.
\end{equation}
Then the Schr\"odinger equation (1) holds with the Hamiltonian
\begin{equation}
H=\frac{1}{2} g^{-\frac{1}{2}}p_{j}g^{\frac{1}{2}}g^{jk}p_{k}+U,
\end{equation}
where $g=\det[g_{jk}]$ , $g^{jk}$ are the matrix elements of the
inverse matrix and
\begin{equation} p_{j}=g_{jk}\frac{dy^{k}}{ds}\end{equation}is the momentum  in
classical mechanics . In quantum mechanics
\begin{equation}
p_{j}=-i\hbar\frac{\partial}{\partial y^{j}}.
\end{equation}
 The corresponding stochastic equation (an analog of eq.(8)) reads
\begin{equation}
dq^{j}=\frac{i\hbar}{m} g^{jk}\partial_{k}\ln(\psi^{g}_{t-s})ds
+\sqrt{\frac{i\hbar}{m}}e^{j}_{k}\circ db_{s}^{k},
\end{equation}
where $e^{j}_{l}e^{k}_{l}=g^{jk}$.

We consider as a simple example
$\tilde{U}=\frac{m\omega^{2}x^{2}}{2}$. Then the ground state
solution of eq.(2) is
\begin{equation}
\psi_{g}(x)=\Big(\frac{\pi
\hbar}{m\omega}\Big)^{-\frac{1}{4}}\exp(-\frac{m\omega}{2\hbar}x^{2}).
\end{equation}
The stochastic equation (8) reads
\begin{equation}
dq=-i\omega qdt+\sqrt{\frac{i\hbar}{m}} db.
\end{equation}
 A simple calculation gives
\begin{equation}
\int dx\vert \psi_{t}^{g}(x)\vert^{2}E[q_{t}(x)q_{t^{\prime}}(x)]
=\frac{\hbar}{2m\omega}\exp(-i\omega\vert t-t^{\prime}\vert).
\end{equation}
The rhs of eq.(26) is the expectation value in the ground state of
the time-ordered products of Heisenberg picture position operators
of the harmonic oscillator. In this sense the stochastic process
$q_{t}$ has a  physical meaning as its correlation functions
coincide with quantum expectation values.

\section{The relativistic quantum field
theory} In the canonical field theory with the Hamiltonian
\begin{equation}{\cal H}=\frac{1}{2}\int d{\bf x}\Big( \Pi^{2}+(\omega
\Phi)^{2}\Big)+\int d{\bf x}V(\Phi),
\end{equation}where $\Pi({\bf x})$ is the canonical momentum
\begin{displaymath}
 \omega=\sqrt{-\triangle+m^{2}}
 \end{displaymath}
and
\begin{displaymath}
[\Phi({\bf x}),\Pi({\bf y})]=i\hbar \delta({\bf x}-{\bf y})
\end{displaymath}
consider the Schr\"odinger equation
\begin{equation}
i\hbar\partial_{t}\psi={\cal H}\psi
\end{equation}
and its imaginary time version
\begin{equation}
-\hbar\partial_{t}\psi={\cal H}\psi.
\end{equation}
Let
\begin{equation}
\psi_{0}=\psi_{g}\chi_{0},
\end{equation}
where $\psi^{g}$ is the ground state. Then, $\chi$ satisfies the
equation (an infinite dimensional version of eq.(6))
\begin{equation} \hbar\partial_{t}\chi=\frac{1}{2}\int d{\bf
x}\Big(- \Pi^{2}-2(\Pi\ln\psi_{g})\Pi\Big)\chi,
\end{equation}where
\begin{equation}
\Pi({\bf x})=-i\hbar\frac{\delta}{\delta \Phi({\bf x})}.
\end{equation}
Eq.(31) is a diffusion equation in infinite dimensional spaces. An
approach to Euclidean field theory based on this equation has been
developed in \cite{albeverio}. It follows that the solution of
eq.(31) can be expressed as
\begin{equation}
\chi_{t}(\Phi)=E\Big[\chi_{0}\Big(\Phi_{t}(\Phi)\Big)\Big],
\end{equation}
where $\Phi_{t}(\Phi)$ is the solution of the stochastic equation
\begin{equation}
d\Phi_{t}({\bf x})=\hbar\frac{\delta}{\delta \Phi({\bf
x})}\ln\psi_{g}dt+\sqrt{\hbar}dW_{t}({\bf x})
\end{equation}with the initial condition $\Phi$. $E[...]$ denotes an expectation value with respect to the
Wiener process (Brownian motion) defined by the covariance
\begin{equation}
E\Big[W_{t}({\bf x})W_{s}({\bf y})\Big]=min(t,s)\delta({\bf
x}-{\bf y}).
\end{equation}
The correlation functions of the Euclidean field in the ground
state $\psi_{g}$ can be expressed by the correlation functions of
the stochastic process $\Phi_{t}$.

 Let us consider the simplest
example:the free field. Then, the ground state is
\begin{equation}
\psi_{g}=\Big(\det(\frac{\pi\hbar}{\omega})\Big)^{-\frac{1}{4}}\exp(-\frac{1}{2\hbar}\Phi\omega\Phi).
\end{equation}
 Eq.(34) reads
\begin{equation}
d\Phi=-\omega\Phi dt+\sqrt{\hbar}dW
\end{equation}
with the solution (with the initial condition $\Phi$ at $t_{0}$)
\begin{equation}
\Phi_{t}=\exp(-\omega (t-t_{0}))\Phi+\sqrt{\hbar}\int_{t_{0}}^{t}
\exp(-\omega(t-s))dW_{s}.\end{equation} We can calculate
\begin{equation}\begin{array}{l}
\int {\cal D}\Phi \vert \psi_{g}(\Phi)\vert^{2}E[\Phi_{t}({\bf
x})\Phi_{t^{\prime}}({\bf x}^{\prime})]\cr =\hbar
\Big(\frac{1}{2\omega}\exp(-\vert t-t^{\prime}\vert
\omega)\Big)({\bf x},{\bf x}^{\prime}).\end{array}
\end{equation}
In eq.(39) the rhs denotes the kernel of the operator.

 If the ground
state has some analycity property  then we can extend the time in
eqs.(28)-(38) $t\rightarrow it$ from imaginary time to the real
time \cite{habajpa} \cite{hababook}. An analytic continuation to
real time of eq.(37) reads

\begin{equation}
d\Phi_{t}=-i\omega\Phi_{t} dt+\sqrt{i\hbar}dW.
\end{equation} The solution is
\begin{equation}
\Phi_{t}(\Phi)=\exp(-i\omega
(t-t_{0}))\Phi+\sqrt{i\hbar}\int_{t_{0}}^{t}
\exp(-i\omega(t-s))dW_{s}.\end{equation} The solution of the
Schr\"odinger equation (28)  has the same  form (33). We can see
that

\begin{equation}\begin{array}{l}
\Big(\psi_{g}, T\Big(\exp(\int
dt\Phi_{t}^{q}(f_{t})\Big)\psi_{g}\Big)\cr= \int d\Phi
\exp(-\frac{1}{\hbar}\Phi\omega\Phi)E\Big[\exp(\int dt d{\bf
x}f_{t}({\bf x})\Phi_{t}({\bf x}))\Big] \cr =
\exp\Big(\frac{\hbar}{2}\int dtdt^{\prime}
\Big(f_{t},(2\omega)^{-1}\exp(-i\omega \vert
t-t^{\prime}\vert)f_{t^{\prime}}\Big)\Big),
\end{array}\end{equation}
where $\Phi_{t}^{q}(f)=\int d{\bf x}\Phi_{t}^{q}({\bf x})f({\bf
x})$ is the quantum field and $T$ denotes the time-ordered
product.

We can transform the complex equation (40) into two real equations
defining
\begin{equation}
\Phi=\phi_{1}+i\phi_{2}
\end{equation}
and \begin{equation} \phi_{+}=\phi_{1}+\phi_{2}
\end{equation}\begin{equation} \phi_{-}=\phi_{1}-\phi_{2}.
\end{equation}Then
it follows from eq.(40) that
\begin{equation}
\phi_{+}=\omega^{-1}\phi_{-}
\end{equation}
and $\phi_{-}$ satisfies the random wave equation
\begin{equation}
\partial_{t}^{2}\phi_{-}=-\omega^{2}\phi_{-}+\sqrt{2\hbar}\omega\partial_{t}W,
\end{equation}
which has the solution
\begin{equation}\begin{array}{l}
\phi_{-}(t)=\cos(\omega t)\phi_{-}(0)+\omega^{-1}\sin(\omega
t)\partial_{t}\phi_{-}(0) \cr
+\sqrt{2\hbar}\int_{0}^{t}\sin(\omega(t-s))dW_{s},
\end{array}\end{equation}
where $\partial_{t}\phi_{-}(0)=\omega\phi_{+}(0)$.

We further develop the theory of stochastic wave equations for the
quantum theory of fields in expanding universes in
\cite{habastoch} as an extension of Starobinsky stochastic
inflation \cite{star}.
\section{Time-dependent reference state}
There are  time-dependent solutions in the Gaussian form of the
Schr\"odinger equation for the free field theory
\begin{equation}
\psi^{g}_{t}(\Phi)=A(t)\exp\Big(\frac{i}{2\hbar}(\Phi\Gamma(t)\Phi+2J_{t}\Phi)\Big).
\end{equation}
In the time-dependent case eq.(8) in quantum field theory  takes
the form \cite{freidlin}
\begin{equation}
d\Phi_{s}=-\Gamma(t-s)\Phi_{s}ds-J_{t-s}ds+\sqrt{i\hbar}dW_{s}.
\end{equation}Let $\Phi_{s}(\Phi)$ be the solution of eq.(50) with
the initial condition $\Phi$ then the solution of the
Schr\"odinger equation (28) with the initial condition
$\psi^{g}_{0}\chi$ is
\begin{equation}
\psi_{t}=\psi^{g}_{t}E\Big[\chi\Big(\Phi_{t}(\Phi)\Big)\Big].
\end{equation}In the model (27) with a time-dependent interaction
$V_{t}$ it would be difficult to find any explicit solution
$\psi_{t}^{g}$. However, we can use the solution $\psi_{t}^{g}$ of
the free field theory and take interaction into account by means
of the Feynman-Kac formula. According to eq.(11) the solution of
the Schr\"odinger equation with the interaction $V_{t} $ reads
\begin{equation}
\psi_{t}(\Phi)=\psi^{g}_{t}E\Big[\exp\Big(-\frac{i}{\hbar}\int_{0}^{t}V_{t-s}(\Phi_{s})ds\Big)\chi\Big(\Phi_{t}(\Phi)\Big)\Big].
\end{equation}
We shall apply eq.(52) for $V$ which is linear in $\Phi$ ($\Phi$
could be the gravitational potential in de Donder gauge as  in
\cite{zurek}; it will have many components when describing
gravitons in the transverse-traceless gauge in secs.7-8).

As an application to the particle motion in a (quantized squeezed)
gravitational wave let us consider the Hamiltonian of a harmonic
oscillator($\tilde{U}=\frac{\omega^{2}x^{2}}{2}$ in eq.(2)
corresponding to a single mode of the gravitational field)
\begin{equation}
H=-\frac{\hbar^{2}}{2}\nabla^{2}+\frac{1}{2}\omega^{2}x^{2}.
\end{equation}
Let us look for a time-dependent solution of the  Schr\"odinger
equation (2) in the form\begin{equation}
\psi^{g}_{t}=A(t)\exp\Big(\frac{i}{2\hbar}(x\Gamma(t)x+2J_{t}x)\Big).
\end{equation}
Then, $\psi^{g}$ is a solution of the Schr\"odinger equation (2)
if
\begin{equation}
i\hbar\partial_{t}\ln A=-\frac{i\hbar}{2}\Gamma+\frac{1}{2}J^{2},
\end{equation}
\begin{equation}
-\partial_{t}J=\Gamma J
\end{equation}
and
\begin{equation}
\partial_{t}\Gamma+\Gamma^{2}+\omega^{2}=0.
\end{equation}
It can be seen that the Riccatti equation (57) is equivalent to
\begin{equation}
\frac{d^{2}u}{dt^{2}}+\omega^{2}u=0,
\end{equation}
where
\begin{equation}
u=\exp(\int^{t}\Gamma).
\end{equation}i.e.$ \Gamma=u^{-1}\partial_{t}u$.
Then, $J_{t}=J_{0}u_{0}u_{t}^{-1}$. The general solution of
eq.(58) is
\begin{equation}
u= \sigma\cos(\omega t)+\delta\sin(\omega t),
\end{equation}
where $\sigma,\delta$ are complex numbers.

Eq.(8)  reads
\begin{equation}\begin{array}{l}
dq_{s}=-\partial_{t}\ln u_{t-s}
q_{s}ds-J_{t-s}ds+\sqrt{i\hbar}db_{s}
\end{array}
\end{equation}
with the solution
\begin{equation}\begin{array}{l}
q_{s}(q)=\frac{u_{t-s}}{u_{t}}q+\sqrt{i\hbar}u_{t-s}\int_{0}^{s}u_{t-\tau}^{-1}db_{\tau}
-u_{0}u_{t-s}J_{0}\int_{0}^{s}u_{t-\tau}^{-2}d\tau.
\end{array}
\end{equation}
We have (as $E[(\int f_{s}db_{s})^{2}]=\int f_{s}^{2}ds$)
\begin{equation}\begin{array}{l}
E[(q_{s}-<q_{s}>)(q_{s^{\prime}}-<q_{s^{\prime}}>)]\cr =i\hbar
u_{t-s}u_{t-s^{\prime}}\int_{0}^{min(s,s^{\prime})}u_{t-\tau}^{-2}d\tau
\cr=(-i\hbar)\omega^{-2}u_{t-s}u_{t-s^{\prime}}\cr
(\sigma^{2}+\delta^{2})^{-1}
\Big(\Gamma(t)-\Gamma(t-min(s,s^{\prime}))\Big).\end{array}
\end{equation}
 If
$\delta a=i\sigma$ then
 $\Gamma(0)=ia^{-1}\omega$ and the solution $\psi_{t}$ starts from a real $i\Gamma$.
 Choosing $J_{0}=a^{-1}\omega x_{0}+ip$ we obtain the squeezed state
 \begin{displaymath}
\psi_{0}^{g}=\exp(-\frac{\omega
(x-x_{0})^{2}}{2a\hbar}+\frac{i}{\hbar}px)
\end{displaymath} with the squeezing $a$ of the coordinate (for a time evolution of
squeezed states see \cite{iwo}).
  The case
$\delta=i\sigma$ corresponds to the ground state (24) ($m=1$). It
can be shown that the formula (63) is continuous with respect to
the limit $\delta\rightarrow i\sigma$.
\section{Linear coupling to an oscillator environment}
Let us consider a model of a system with a Lagrangian ${\cal
L}_{\xi}$ described by a coordinate $\xi$ linearly interacting
with an oscillator. We have the Lagrangian
\begin{equation}
{\cal L}={\cal L}_{\xi}+
\frac{1}{2}((\frac{dq}{ds})^{2}-\omega^{2}q^{2})+ qf_{s}(\xi).
\end{equation}
In the model with a linear coupling and the quadratic Lagrangian
for the $q$ variable (64) the functional integral in eq.(11) can
be reduced to the Gaussian integral
\begin{equation}\begin{array}{l} \chi_{t}(x) =\int
dk\chi_{0}(k)E\Big[\exp\Big(\frac{i}{\hbar}\int_{0}^{t}dsq_{s}(x)f_{t-s}(\xi_{s})+ikq_{t}(x)\Big)\Big]
\end{array}\end{equation}
where we used the Fourier representation of $\chi_{0}(x)$ (we use
the same notation for a function and its Fourier transform)
\begin{displaymath}
\chi_{0}(x)=\int dk\chi_{0}(k)\exp(ikx)
\end{displaymath}
For a Gaussian variable $q_{s}$ we have (for any number
$\alpha_{s}$)
\begin{equation}\begin{array}{l}
E[\exp(\alpha_{s}q_{s})]\cr=\exp\Big(\alpha_{s}<q_{s}>+\frac{1}{2}<(\alpha_{s}q_{s}-\alpha_{s}<q_{s}>)^{2}>\Big)
\end{array} \end{equation} The formula (66) can easily be generalized to
$\int ds\alpha_{s}q_{s}$ with $q_{s}$ of eq.(62)) and to an
infinite number of modes $q$ (as will be done in secs.7-8).

 We are interested in the scattering
amplitude from an initial state
$\psi_{0}^{g}(x)\chi_{i}(x)\phi_{i}(\xi)$ to the final state

$\psi_{0}^{g}(x)\chi_{f}(x)\phi_{f}(\xi)$. We apply the
transformation of sec.2 only to the oscillator path integral.
Then, according to eq.(11)  the amplitude $a_{fi}$ is (where
$(\psi_{0}^{g}\chi_{i}(x)\phi_{i})_{t}$ means the unitary
evolution of the wave function)
\begin{equation}\begin{array}{l}
a_{fi}=(\psi_{0}^{g}\chi_{f}\phi_{f},(\psi_{0}^{g}\chi_{i}(x)\phi_{i})_{t})\cr
=\int dxd\xi\int{\cal D}\xi\exp(\frac{i}{\hbar}\int ds {\cal
L}_{\xi})
\overline{\psi_{0}^{g}(x)}\overline{\chi_{f}(x)}\overline{\phi_{f}(\xi)}\cr
\psi_{t}^{g}
\phi_{i}(\xi_{t}(\xi))E\Big[\exp(\frac{i}{\hbar}\int_{0}^{t}q_{s}f_{t-s}(\xi_{s}))\chi_{i}(q_{t}(x))\Big],
\end{array}
\end{equation}where the initial state of the oscillator is $\psi_{0}^{g}\chi_{i}$.

 If we do not observe the final
states of the oscillator and average the probability $ P(i,f)$ of
the transition $\phi_{i}\rightarrow \phi_{f}$ over these states
using the completeness relation
\begin{equation}
\sum_{f}\overline{\psi_{t}^{g}(x)\chi_{f}(x)}\psi_{t}^{g}(x^{\prime})\chi_{f}(x^{\prime})=\delta(x-x^{\prime})
\end{equation}
then we obtain
\begin{equation}\begin{array}{l}
P(i,f)=\sum_{f} \vert a_{fi}\vert^{2}=\int dxd\xi d\xi^{\prime}
{\cal D}\xi{\cal D}\xi^{\prime}\vert\psi_{t}^{g}(x)\vert^{2}\cr
\phi_{f}(\xi)\overline{\phi_{f}}(\xi^{\prime})
\overline{\phi_{i}}(\xi_{t}(\xi))\phi_{i}(\xi_{t}(\xi^{\prime}))\cr\exp(-\frac{i}{\hbar}\int
ds {\cal L}_{\xi}+\frac{i}{\hbar}\int ds {\cal
L}_{\xi^{\prime}})\cr
E\Big[\exp\Big(\frac{i}{\hbar}\int_{0}^{t}q_{s}f_{t-s}(\xi_{s}^{\prime})\Big)\chi_{i}(q_{t}(x))\Big)\cr
\exp\Big(-\frac{i}{\hbar}\int_{0}^{t}q^{*}_{s}f_{t-s}(\xi_{s})\Big)\overline{\chi_{i}}(q^{*}_{t}(x))\Big].
\end{array}\end{equation}
where $q^{*}$ denotes a complex conjugation of an independent
version of the process $q_{t}$ and
$\xi_{s}^{\prime}(\xi^{\prime})$ is another realization of the
path $\xi_{s}$.

If we define the density matrix $\rho$ as an average over the
environment of the oscillator then the density matrix of the $\xi$
system is
\begin{equation}\begin{array}{l}
\rho_{t}(\xi,\xi^{\prime})=\int dx {\cal D}\xi{\cal
D}\xi^{\prime}\vert\psi_{t}^{g}(x)\vert^{2}\exp(-\frac{i}{\hbar}\int
ds {\cal L}_{\xi}+\frac{i}{\hbar}\int ds {\cal
L}_{\xi^{\prime}})\cr\overline{\phi_{i}}(\xi_{t}(\xi))\phi_{i}(\xi_{t}(\xi^{\prime}))
E\Big[\exp(\frac{i}{\hbar}\int_{0}^{t}q_{s}f_{t-s}(\xi_{s}^{\prime})\Big)\chi_{i}(q_{t}(x))\cr
\exp(-\frac{i}{\hbar}\int_{0}^{t}q^{*}_{s}f_{t-s}(\xi_{s})\Big)\overline{\chi_{i}}(q^{*}_{t}(x))\Big],
\end{array}\end{equation}
It can be seen from eqs.(69)-(70) that the calculations of the
density matrix and the transition probability are closely related.
 When calculating the transition probability we need to perform
an extra $(\xi,\xi^{\prime})$  integral over the final states
$\overline{\phi_{f}(\xi)}\phi_{f}(\xi^{\prime})$ of the $\xi$
system in comparison to the calculation of the density matrix in
eq.(70).

If $\chi_{i}=1$ then according to eq.(66) the expectation value
(67) is
\begin{equation}\begin{array}{l}
E\Big[\exp\Big(\frac{i}{\hbar}\int_{0}^{t}q_{s}f_{t-s}(\xi_{s})\Big)\Big]
\cr=\exp\Big(\frac{i}{\hbar}\int_{0}^{t}<q_{s}>f_{t-s}(\xi_{s})\Big)
\cr\exp\Big(-\frac{1}{2\hbar^{2}}\int_{0}^{t}dsds^{\prime}\cr
E[(q_{s}-<q_{s}>)(q_{s^{\prime}}-<q_{s^{\prime}}>)]
f_{t-s}(\xi_{s})f_{t-s^{\prime}}(\xi_{s^{\prime}})\Big).\end{array}\end{equation}

\section{Particle
interacting with gravitons:PWZ model in one mode approximation}
Parikh,Wilczek and Zahariade \cite{wilczek} consider two masses $M
$ and $m_{0}$ interacting with gravitational field $q_{\omega}$.
In a one mode approximation the Lagrangian describing the geodesic
deviation of the $m_{0}$ mass ( in the free falling frame and
$M>>m_{0}$) is \cite{wilczek}\cite{soda}
\begin{equation}\begin{array}{l} {\cal
L}=\frac{1}{2}((\frac{dq_{\omega}}{ds})^{2}-\omega^{2}q_{\omega}^{2})+\frac{1}{2}m_{0}(\frac{d\xi}{ds})^{2}-
m_{0}\lambda\frac{dq_{\omega}}{ds}\frac{d\xi}{ds}\xi-V(\xi)
\cr=\frac{1}{2}g_{jk}\frac{dy^{j}}{ds}\frac{dy^{k}}{ds}-\frac{1}{2}\omega^{2}q_{\omega}^{2}-V(\xi),\end{array}\end{equation}
where $\lambda=\sqrt{8\pi G}$, $y=(q_{\omega},\xi)$, $\xi$ is the
geodesic deviation (the potential $V(\xi)$ is absent in
\cite{wilczek} but we add it here for further applications in the
next sections).
\begin{equation}
(g_{jk})=\left[\begin{array}{ccc}1 &-m_{0}\lambda\xi\\
-m_{0}\lambda\xi& m_{0}
\end{array}\right].\end{equation}
The Lagrange equation is \begin{displaymath}
\frac{d^{2}\xi}{ds^{2}}=m_{0}\lambda
\frac{d^{2}q_{\omega}}{ds^{2}}\xi.\end{displaymath} The
Hamiltonian is determined by eq.(20) with
\begin{equation}
[g^{jk}]=(m_{0}-\lambda^{2}\xi^{2})^{-1}\left[\begin{array}{ccc}m_{0} &m_{0}\lambda\xi\\
m_{0}\lambda\xi& 1
\end{array}\right].\end{equation}
We could  quantize the model (72) with the explicit quantum
version (20) of the Hamiltonian. However, the averaging over
gravitons is simple only in the influence functional approach.

 In the PWZ model (72), where the gravitational field is described by one mode $q_{\omega}$, when
   we apply the transformation (25) to $q_{s}$ with the ground state
(24) ($m=1$) then from eq.(70) we obtain
\begin{equation}
\begin{array}{l}
\rho_{t}(\xi,\xi^{\prime})=\int dx
\exp(-\frac{\omega}{\hbar}x^{2})\cr
E\Big[\exp\Big(\frac{i}{\hbar}\int_{0}^{t}( \frac{m_{0}}{2}
\frac{d\xi}{ds}\frac{d\xi}{ds}-\frac{m_{0}}{2} \frac{d\xi^{\prime
}}{ds}\frac{d\xi^{\prime }}{ds}-V(\xi)+V(\xi^{\prime})
\Big)\cr\exp\Big(-\frac{i\lambda m_{0}}{2\hbar}\int_{0}^{t}
(q_{t-s}(x)\frac{d^{2}\xi^{2}}{ds^{2}} -
q_{t-s}^{*}(x)\frac{d^{2}\xi^{\prime
2}}{ds^{2}})\Big)\cr\chi_{i}(q_{t}(x))
\overline{\chi_{i}}(q^{*}_{t}(x))\overline{\phi_{i}}(\xi_{t}(\xi))\phi_{i}(\xi_{t}(\xi^{\prime}))\Big]\cr
\equiv K_{t}\rho_{0},
\end{array}\end{equation}where $q^{*}$ is another realization
of the process $q$ (and the complex conjugation of this
realization). We have changed $s\rightarrow t-s$ in the integral
in the exponential and inserted
$f_{s}=\frac{m_{0}}{2}\frac{d^{2}\xi^{2}}{ds^{2}}$. We have
denoted the evolution kernel of the density matrix by $K_{t}$ .

The calculation of the expectation value (75) according to eq.(71)
gives (we use the solution (41), where $\Phi\rightarrow q$, of
eq.(25) and assume that the initial condition $\chi_{i}=1$ )
\begin{equation}\begin{array}{l}\rho_{t}\simeq\int dx{\cal D}\xi{\cal D}\xi^{\prime}\exp(-\frac{\omega x^{2}}{\hbar})
\cr E\Big[\exp\Big(\frac{i}{\hbar}\int_{0}^{t}(
\frac{m_{0}}{2}\frac{d\xi}{ds}\frac{d\xi}{ds}-\frac{m_{0}}{2}
\frac{d\xi^{\prime }}{ds}\frac{d\xi^{\prime }}{ds}\cr
-V(\xi)+V(\xi^{\prime}) -\frac{i\lambda}{\hbar}
\int_{0}^{t}(q_{t-s}f_{s}-q^{*}_{t-s}f^{\prime}_{s})ds\Big)\Big]\cr
\simeq\int dx{\cal D}\xi{\cal D}\xi^{\prime}\exp(-\frac{\omega
x^{2}}{\hbar}) \exp\Big(-\frac{i\lambda}{\hbar}
\int_{0}^{t}(q\exp(-i\omega (t-s))f_{s}\cr-q\exp(i\omega
(t-s))f^{\prime}_{s})ds\Big)
\cr\exp\Big(-\frac{\lambda^{2}}{2\hbar^{2}}\int_{0}^{t}dsds^{\prime}\cr

\Big(E[(q_{t-s}-<q_{t-s}>)(q_{t-s^{\prime}}-<q_{t-s^{\prime}}>)]f_{s}f_{s^{\prime}}\cr
+E[(q^{*}_{t-s}-<q^{*}_{t-s}>)(q^{*}_{t-s^{\prime}}-<q^{*}_{t-s^{\prime}}>)]f^{\prime}_{s}f^{\prime}_{s^{\prime}}\Big)\Big),
\end{array}
\end{equation} where
$f=\frac{m_{0}}{2}\frac{d^{2}\xi^{2}}{ds^{2}},f^{\prime}=\frac{m_{0}}{2}\frac{d^{2}\xi^{\prime
2}}{ds^{2}}$. In eq.(76) we have (this is the special case of
eq.(63) with $u_{s}=\exp(i\omega s)$)
\begin{equation}\begin{array}{l}
E[(q_{t-s}-<q_{t-s}>)(q_{t-s^{\prime}}-<q_{t-s^{\prime}}>)]\cr=\frac{\hbar}{2\omega}\Big(\exp(-i\omega\vert
s-s^{\prime}\vert)
-\exp(-i\omega(2t-s-s^{\prime}))\Big).\end{array}
\end{equation}
If the oscillator is in a time-dependent state then  we should
insert the solution (62) in the Feynman formula (75). Hence,
instead of eq.(76) we have (for typographical reasons from now on
we identify $\overline{\Gamma}=\Gamma^{*}$, when acting on $q_{t}$
the star $*$ has an extra meaning:it means complex conjugation and
an independent realization of the process $q_{t}$)
\begin{equation}
\begin{array}{l}\int dx\vert\exp(i\frac{\Gamma(t)
x^{2}}{2\hbar})\vert^{2} E\Big[\exp\Big(\frac{i\lambda}{\hbar}
\int_{0}^{t}(q_{s}f_{t-s}-q^{*}_{s}f^{\prime}_{t-s})ds\Big)\Big]\cr
=\int dx\exp(i\frac{\Gamma(t)
x^{2}}{2\hbar})\exp(-i\frac{\Gamma^{*}(t) x^{2}}{2\hbar})\cr
 \exp\Big(\frac{-i\lambda}{\hbar}
\int_{0}^{t}(<q_{t-s}>f_{s}-<q_{t-s}^{*}>f^{\prime}_{s})ds\Big)
\cr\exp\Big(-\frac{\lambda^{2}}{2\hbar^{2}}\int_{0}^{t}
dsds^{\prime}
\cr\Big(E[(q_{t-s}-<q_{t-s}>)(q_{t-s^{\prime}}-<q_{t-s^{\prime}}>)]f_{s}f_{s^{\prime}}\cr
+E[(q^{*}_{t-s}-<q^{*}_{t-s}>)(q^{*}_{t-s^{\prime}}-<q^{*}_{t-s^{\prime}}>)]
f^{\prime}_{s}f^{\prime}_{s^{\prime}}\Big)\Big),
\end{array}
\end{equation}
where
\begin{equation}\begin{array}{l}
E[(q_{t-s}-<q_{t-s}>)(q_{t-s^{\prime}}-<q_{t-s^{\prime}}>)]\cr=i\hbar
u_{s}u_{s^{\prime}}\int_{0}^{min(t-s,t-s^{\prime})}d\tau
u(t-\tau)^{-2}\cr=-i\hbar
\omega^{-2}u_{s}u_{s^{\prime}}(\sigma^{2}+\delta^{2})^{-1}\cr(\Gamma(t)-\Gamma(max(s,s^{\prime}))).
\end{array}\end{equation}
After calculation of the $x$ integral in eqs.(76) and (78) we
obtain a quadratic functional of $f_{s}$ and
$f^{\prime}_{s^{\prime}}$ in the exponential. In the simplest case
(76) of the ground state of the oscillator we
obtain\begin{equation}\begin{array}{l} \rho_{t}\simeq\int {\cal
D}\xi{\cal D}\xi^{\prime}\cr \exp\Big(\frac{i}{\hbar}\int_{0}^{t}(
\frac{m_{0}}{2} \frac{d\xi}{ds}\frac{d\xi}{ds}-\frac{m_{0}}{2}
\frac{d\xi^{\prime }}{ds}\frac{d\xi^{\prime
}}{ds}-V(\xi)+V(\xi^{\prime}) )\Big)\cr
\exp\Big(-\frac{\lambda^{2}}{4\hbar\omega}\int_{0}^{t}
dsds^{\prime}\Big(f_{s}f_{s^{\prime}}\exp(-i\omega\vert
s-s^{\prime}\vert)\cr+f^{\prime}_{s}f^{\prime}_{s^{\prime}}\exp(i\omega\vert
s-s^{\prime}\vert)-2f_{s^{\prime}}f^{\prime}_{s}\exp(i\omega(s-s^{\prime})\Big)\Big).
\end{array}\end{equation}We write \begin{equation} X=\frac{1}{2}(\xi+\xi^{\prime })
\end{equation}
\begin{displaymath}
y=\xi-\xi^{\prime }.
\end{displaymath}We expand the exponential in eq.(80) in $y$. Then,  the terms
 independent of $y$ cancel and there remains (till  the terms
quadratic in $y$ )\begin{equation}\begin{array}{l}\rho_{t}\simeq
\int{\cal D}X{\cal D}y
\exp\Big(\frac{i}{\hbar}\int_{0}^{t}y(m_{0}\frac{d^{2}X}{ds^{2}}-V^{\prime}(X))\Big)\cr\exp\Big(-\frac{\lambda^{2}m_{0}^{2}}{8\hbar\omega}\int_{0}^{t}
dsds^{\prime}\Big(-i\sin(\omega(s-s^{\prime}))(\frac{d^{2}Xy}{ds^{\prime
2}}\frac{d^{2}X^{2}}{ds^{2}}
\cr-\frac{d^{2}Xy}{ds^{2}}\frac{d^{2}X^{2}}{ds^{\prime
2}})+\frac{d^{2}Xy}{ds^{2}}\frac{d^{2}Xy}{ds^{\prime
2}}\cos(\omega(s-s^{\prime}))\Big)\Big)\rho_{0}(X_{t},y_{t}).
\end{array}\end{equation}
The term linear in $y$ gives a modification of the equation of
motion of the $\xi$ coordinate whereas the term quadratic in $y$
is a noise acting upon the particle \cite{hk}.

In the expression (78) of the time-dependent reference state we
obtain
\begin{equation}\begin{array}{l} \rho_{t}\simeq
\exp\Big(-\frac{i}{2\hbar}(\Gamma(t)-\Gamma^{*}(t))^{-1}\cr\times\Big(\lambda\int_{0}^{t}(u_{t}^{-1}u_{s}f_{s}
 -u_{t}^{*-1}u_{s}^{*}f_{s}^{\prime})ds\Big)^{2}\cr
 -\frac{i\lambda^{2}}{2\hbar\omega^{2}}\int_{0}^{t}\Big(u_{s}u_{s^{\prime}}(\sigma^{2}+\delta^{2})^{-1}
(\Gamma(t)-\Gamma(max(s,s^{\prime})))f_{s}f_{s^{\prime}}\cr
-u^{*}_{s}u^{*}_{s^{\prime}}(\sigma^{* 2}+\delta^{* 2})^{-1}
(\Gamma^{*}(t)-\Gamma^{*}(max(s,s^{\prime})))f^{\prime}_{s}f^{\prime}_{s^{\prime}}
\Big)dsds^{\prime}\Big).\end{array}
\end{equation}
We  expand (83) in $y$ again. Let us consider the general
expression appearing after an expansion in $y$ till the second
order terms in eqs.(82)-(83)
\begin{displaymath}\begin{array}{l}
\int {\cal D}X {\cal D}y\cr
\exp\Big(\frac{i}{\hbar}\int_{0}^{t}y(m_{0}\frac{d^{2}X}{ds^{2}}+V^{\prime}(X)
+L(X)+\frac{i}{2\hbar}My)\Big)\rho_{0}(X,y)\cr\equiv\int {\cal D}X
{\cal
D}y\exp\Big(\frac{i}{\hbar}\int_{0}^{t}(y\tilde{L}+\frac{i}{2\hbar}yMy)\Big)\rho_{0}(X,y)\cr
=\int {\cal D}X {\cal D}y \exp\Big(-\frac{1}{2\hbar^{2}}(y-i\hbar
M^{-1}\tilde{L})M(y-i\hbar M^{-1}\tilde{L})\cr
-\frac{1}{2}\tilde{L}M^{-1}\tilde{L}\Big)
\rho_{0}(X,y),\end{array}\end{displaymath}where by $L$ we denote a
functional of $X$, $M$ is an operator and by $\tilde{L}$ we denote
the term proportional to $y$. If we introduce $\tilde{X}=
M^{-\frac{1}{2}}\tilde{L}$ then $\tilde{X}$ becomes a Gaussian
variable with the white noise distribution which can be
represented as $\partial_{s}b_{s}$. If $\rho_{0}$ depends only on
$X$ then the factor depending on $y$ is integrated out
contributing just a constant. The calculation of $\rho_{t}$ is
reduced to an average over solutions of the stochastic equation
\begin{equation}
m_{0}\frac{d^{2}X}{ds^{2}}+V^{\prime}(X)+L(X)=M^{\frac{1}{2}}\partial_{s}b_{s}.
\end{equation}
In general, there still will be the Gaussian integral over $y$ so
that the expression for the density matrix can be obtained in the
form of an expectation value over the solutions of the stochastic
equation (84) and the $y$ terms resulting from an expansion in $y$
of $\rho_{0}(X,y)$ (this is an expansion in $\hbar$).

 In the next
section dealing with a thermal state of the gravitational field we
approximate $L$ and $M$ by local functions of $X$.

\section{Infinite number of modes:thermal state of gravitons}
We are to generalize the results of sec.6 to an infinite number of
modes of the gravitational field. We could do it in the
formulation of sec.6 by means of a representation of the thermal
state as a sum over eigenstates with a proper weight factor. We
have studied earlier \cite{hk} an analogous model of a particle
geodesic motion in an environment of quantum gravitational waves
in a thermal state. There are minor changes from the setting of
Parikh, Wilczek and Zahariade \cite{wilczek}where the effect of
gravitons on geodesic deviation is considered.The Lagrangian (72)
with an infinite number of modes  in the coordinate space is
\cite{soda}
\begin{equation}
\begin{array}{l}
{\cal L}=\frac{1}{4} \int d{\bf x}h_{\alpha}(s,{\bf
x})(-\partial_{s}^{2}+\triangle)h_{\alpha}(s,{\bf
x})+\frac{1}{2}m_{0}\frac{d\xi_{r}}{ds}\frac{d\xi_{r}}{ds}\cr-
\frac{1}{4}m_{0}\lambda
\partial_{s}^{2}h_{rl}^{w}(s,\xi_{s})\xi_{s}^{r}\xi_{s}^{l}-
\frac{1}{4}m_{0}\lambda
\partial_{s}^{2}h_{rl}^{q}(s,\xi_{s})\xi_{s}^{r}\xi_{s}^{l},
\end{array}\end{equation} where the metric perturbation $h_{rl}$ of the Minkowski metric $\eta_{\mu\nu}$
($\eta_{\mu\nu}\rightarrow \eta_{\mu\nu}+h_{\mu\nu}$)
 is in the  transverse-traceless gauge,  the geodesic deviation $\xi$ in a gravitational
 transverse traceless gauge has only  the spatial components $\xi_{r}$.
 We apply the decomposition of $h_{rl}=h_{rl}^{w}+
 h_{rl}^{q}$ into the classical wave solution $h_{rl}^{w}$ and
 the quantized (graviton ) part $ h_{rl}^{q}$. $ h_{rl}^{q}$ is decomposed in the amplitudes
 $h_{\alpha}$ (where
 $\alpha=+,\times$,in the linear polarization) by means of the
 polarization tensors $e^{\alpha}_{rl}$
 ( as $h_{rl}=e^{\alpha}_{rl}h_{\alpha}$) \cite{mag}\cite{arxiv}\cite{gravitation} \begin{displaymath}\begin{array}{l}
 h_{rl}(t,{\bf
 x})=h_{rl}^{w}(t,{\bf x})\cr+(2\pi)^{-\frac{3}{2}}\int d{\bf
k}(h_{\alpha}e^{\alpha}_{rl}\exp(-i{\bf
kx})+h_{\alpha}^{*}e^{\alpha}_{rl}\exp(i{\bf kx})).
\end{array}\end{displaymath}
 With the infinite
number of gravitational wave modes in the model (72) of sec.6 we
shall have $\omega=\vert {\bf k}\vert$ (we set the velocity of
light $c=1$).$ H=H_{+}+H_{\times}$ is the sum of the independent
Hamiltonians (27) with $V=m=0$.

The $h_{rl}^{q}$ are quantized at finite temperature $T$ with the
Gibbs distribution $\exp(-\beta H)$, $\beta^{-1}=k_{B}T$ where
$k_{B}$ is the Boltzman constant. The classical part of the
particle-wave interaction can be considered as a time-dependent
external potential $V(\xi)$ with
\begin{equation}
V=\frac{m_{0}}{4}\partial_{s}^{2}h^{w}_{kn}(s,\xi_{s})\xi_{s}^{k}\xi_{s}^{n}.
\end{equation}
 Denote
\begin{equation}
f^{rl}=\frac{m_{0}}{2}\frac{d^{2}}{ds^{2}}\xi^{r}\xi^{l}
\end{equation}\begin{equation}
f^{\prime rl}=\frac{m_{0}}{2}\frac{d^{2}}{ds^{2}}\xi^{\prime
l}\xi^{\prime r}.
\end{equation}
Then, as in\cite{hk} (there are some misprints of signs in
\cite{hk}; for the derivation of the thermal formula  see
\cite{kleinert},sec.18, see also \cite{hu}\cite{hu2}\cite{prd};the
gravitational case is analogous to the electromagnetic one treated
in \cite{hkem}) we obtain for the density matrix evolution kernel
\begin{equation}\begin{array}{l}
K_{t}(\xi,\xi^{\prime})=\int D\xi
D\xi^{\prime}\exp(\frac{im_{0}}{2\hbar}\int_{0}^{t}ds(\frac{d\xi_{r}}{ds}\frac{d\xi_{r}}{ds}
-\frac{d\xi_{r}^{\prime}}{ds}\frac{d\xi_{r}^{\prime}}{ds})
\cr\exp(\frac{i}{\hbar}\int_{0}^{t}(V_{t-s}({\xi_{s}})-V_{t-s}(\xi_{s}^{\prime}))
\cr\exp\Big(\frac{\lambda^{2}}{\hbar^{2}}\int_{0}^{t}ds\int_{0}^{s}ds^{\prime}\Big((f^{rl}-f^{\prime
rl})C_{rl:mn}(f^{mn}+f^{\prime mn})\cr -(f^{rl}-f^{\prime
rl})A_{rl:mn}(f^{mn}-f^{\prime mn})\Big).\end{array}\end{equation}
In (89) we have a decomposition of the finite temperature
transverse-traceless graviton propagator $D$ into the real and
imaginary parts $D=A+iC$
\begin{equation}\begin{array}{l}
A_{rl;mn}({\bf x}-{\bf x}^{\prime},s-s^{\prime})=2\beta
\hbar(2\pi)^{-3}\int \frac{\bf d{\bf
k}}{2k}\Lambda_{rl;mn}\cr\cos({\bf k}({\bf x}-{\bf x}^{\prime}))
\cos (k(s-s^{\prime}))\coth(\frac{\hbar\beta k}{2}),\end{array}
\end{equation}\begin{equation}\begin{array}{l}
C_{rl;mn}({\bf x}-{\bf
x}^{\prime},s-s^{\prime})=2\hbar(2\pi)^{-3}\int \frac{\bf d{\bf
k}}{2k}\Lambda_{rl;mn}\cr\cos({\bf k}({\bf x}-{\bf x}^{\prime}))
\sin (k(s-s^{\prime})),\end{array}
\end{equation}where
\begin{displaymath}\begin{array}{l}
2\Lambda_{ij;mn}=2e^{\alpha}_{ij}e^{\alpha}_{mn}=(\delta_{im}-k^{-2}k_{i}k_{m})
(\delta_{jn}-k^{-2}k_{j}k_{n})\cr+(\delta_{in}-k^{-2}k_{i}k_{n})
(\delta_{jm}-k^{-2}k_{j}k_{m})\cr-\frac{2}{3}(\delta_{ij}-k^{-2}k_{i}k_{j})
(\delta_{nm}-k^{-2}k_{n}k_{m}).\end{array}
\end{displaymath}  We neglect the dependence on ${\bf x}$ in
eqs.(90)-(91). Then the angular average over ${\bf k}k^{-1}$ gives
\begin{equation}
\frac{1}{4\pi}<\Lambda_{ij;mn}>=\frac{1}{5}(\delta_{im}\delta_{jn}+\delta_{in}\delta_{jm})
-\frac{2}{15}\delta_{ij}\delta_{nm}.
\end{equation} When we  neglect the ${\bf x}$ dependence of the propagators
($\xi^{r}-\xi^{\prime r}$ in eq.(89) should be inserted as $x^{r}-
x^{\prime r}$and we assume that $\xi$ is small in comparison with
the gravitational wave length, hence  it can be set to zero) and
average over the angles then the $k$-integral $d{\bf k}\simeq 4\pi
dkk^{2}$ in the high temperature limit $\beta\hbar \rightarrow 0$
of $A_{rl;mn}$ in eq.(90) gives $\delta(s-s^{\prime})$. In
$C_{rl;mn}$ (91) we write (as in \cite{hk})
$\sin(k(s-s^{\prime}))=-k^{-1}\partial_{s}\cos(k(s-s^{\prime}))$.
Then, integrating over $k$ we obtain
$\partial_{s}\delta(s-s^{\prime})$ . In such a case the evolution
kernel of the density matrix reads

\begin{equation}\begin{array}{l}
K_{t}(\xi;\xi^{\prime})=\int D\xi
D\xi^{\prime}\exp\Big(\frac{im_{0}}{2\hbar}\int_{0}^{t}ds(\frac{d\xi_{r}}{ds}\frac{d\xi_{r}}{ds}\cr
-\frac{d\xi_{r}^{\prime}}{ds}\frac{d\xi_{r}^{\prime}}{ds})
+\frac{i}{\hbar}\int_{0}^{t}(V_{t-s}({\xi_{s}})-V_{t-s}(\xi_{s}^{\prime}))\Big)\cr
\exp\Big(-i\frac{\gamma}{2\hbar}\int_{0}^{t}ds\Big((q^{rl}-q^{\prime
rl})\partial_{s}(q^{rl}+q^{\prime rl})\cr
-\frac{w}{2\hbar^{2}}(q^{rl}-q^{\prime rl})(q^{rl}-q^{\prime
rl})\Big)\Big),\end{array}\end{equation}

where \begin{equation}
q^{rl}=\frac{1}{2}\frac{d^{2}}{ds^{2}}(\xi^{r}\xi^{l}-\frac{1}{3}\delta^{rl}\xi_{j}\xi_{j})
\end{equation}results from a resummation of $f^{rl}$ with
$<\Lambda_{ij;rl}>$
\begin{equation}
\gamma=\frac{8\pi G m_{0}^{2}}{10\pi }
\end{equation}
\begin{equation}
w=\frac{2\gamma}{\beta}
\end{equation}
We expand eq.(93) around $X$ (81) (now with three spatial indices)
\begin{displaymath}
q^{rl}-q^{\prime
rl}=\frac{d^{2}}{ds^{2}}(X^{r}y^{l}+X^{l}y^{r}-\frac{2}{3}\delta^{rl}X^{j}y^{j})
\end{displaymath}
In the exponential (93) the term linear in y becomes
\begin{equation}\begin{array}{l}
y_{n}\Big(-\frac{d^{2}X_{n}}{ds^{2}}+\frac{\lambda}{2}\frac{d^{2}h^{w}_{nr}}{ds^{2}}X_{r}\cr
+\frac{8\pi
Gm_{0}}{10\pi}X_{l}\frac{d^{5}}{ds^{5}}(\frac{1}{3}X_{r}X_{r}\delta_{nl}-X_{n}X_{l})\Big)
\end{array}\end{equation}
The term quadratic in $y$ is the noise term. For low temperature
we obtain in general the non-local and non-Markovian stochastic
equation (84). In the high temperature limit $\beta\hbar
\rightarrow 0$ the calculation of the evolution kernel is reduced
to an expectation value over the solutions of the stochastic
equation
\begin{equation}\begin{array}{l}
-\frac{d^{2}X_{n}}{ds^{2}}+\frac{\lambda}{2}h^{w}_{nr}X_{r}+\frac{8\pi
Gm_{0}}{10\pi}X^{l}\frac{d^{5}}{ds^{5}}(\frac{1}{3}X_{r}X_{r}\delta_{nl}-X_{n}X_{l})\cr
=m_{0}^{-1}\sqrt{w}(M^{\frac{1}{2}})_{nr}\partial_{s}b_{s}^{r}.
\end{array}\end{equation}
As explained in the derivation of eq.(84) the term quadratic in
$y$ defines the operator $M$. From eq.(93) we obtain that $M$ is
an operator defined by the bilinear form (on the rhs of eq.(98) we
have the square root of the matrix M)
\begin{displaymath}\begin{array}{l}
w y^{r}M^{rl}y^{l}=2w\int ds\Big(
\frac{d^{2}}{ds^{2}}(X^{j}y^{l})\frac{d^{2}}{ds^{2}}(X^{j}y^{l})\cr
+\frac{d^{2}}{ds^{2}}(X^{j}y^{l})\frac{d^{2}}{ds^{2}}(X^{l}y^{j})
-\frac{2}{3}\frac{d^{2}}{ds^{2}}(X^{j}y^{j})\frac{d^{2}}{ds^{2}}(X^{l}y^{l})\Big)\cr
= \frac{\lambda^{2}}{4\pi}\beta^{-1}y^{r}X^{k}{\cal
M}_{rk;ln}y^{l}X^{n}\end{array}\end{displaymath} where
\begin{equation}
{\cal
M}_{rk;ln}(s,s^{\prime})=<\Lambda_{rk;ln}>\partial_{s}^{2}\partial_{s^{\prime}}^{2}\delta(s-s^{\prime})
\end{equation}We derive equation (98) with the noise (99) in the Appendix working directly with the classical model of thermal
gravitational waves. The modification of the deviation equation
coincides with the one of ref.\cite{soda} (up to an infinite
renormalization term). It is discussed already in
\cite{gravitation} (sec.36.8) and in more detail in \cite{quinn}
(see in particular the last section of this paper).
  The spectrum of the noise (of the quadrupole $q^{rl}$) as seen from eqs.(89)-(90) is $8\pi G
  k\hbar\coth(\frac{\hbar\beta k}{2})$ which at low
  temperature is  $8\pi G
  k$ and at high temperature $8\pi G \beta^{-1}$ ( the spectrum of
  the coordinates $\xi$, as considered by \cite{soda}
  in eqs.(3.13)-(3.14), is multiplied by $k^{4}$ as a result of the
  fourth order derivative in eq.(99)).
\section{General Gaussian state of the graviton}
We generalize the results of  sec.6 on averaging over the
oscillator modes in a squeezed state to infinite number of modes
of the gravitational field. The gravitational perturbation in the
transverse-traceless gauge is decomposed into the amplitudes
$h^{\alpha}$ by means of the polarization tensors
$e_{rl}^{\alpha}$,$\alpha=+,\times$,
$h_{rl}=e^{\alpha}_{rl}h_{\alpha}$. The gravitational Hamiltonian
$H=H_{+}+H_{\times}$ has the same form as for two independent
scalar fields $\Phi\rightarrow h_{\alpha}$ (eq.(27) with $m=U=0$).
In such a case we have two independent equations (57) for
$\Gamma_{\alpha}$ and two independent $u_{\alpha}$. The Lagrangian
(72)  in multimode case takes the form of a sum over modes ( we
omit the classical part $h^{w}$ and neglect the dependence of
$h^{q}(s,{\bf x})$ on spatial coordinates as in eq.(93) of the
previous section))
\begin{displaymath}
\begin{array}{l}
{\cal L}=\frac{1}{4}\int d{\bf k} h^{*}_{\alpha}(s,{\bf
k})(-\partial_{s}^{2}-k^{2})h_{\alpha}(s,{\bf
k})+\frac{1}{2}m_{0}\frac{d\xi_{r}}{ds}\frac{d\xi_{r}}{ds}\cr-
\frac{1}{4}m_{0}\lambda (2\pi)^{-\frac{3}{2}}\int d{\bf
k}\partial_{s}^{2}h_{rl}^{q}(s,{\bf k})\xi_{r}\xi_{l},
\end{array}\end{displaymath}
The last term will be expressed as
 \begin{displaymath}\begin{array}{l}
\frac{\lambda m_{0}}{2}(2\pi)^{-\frac{3}{2}}\int dsd{\bf
k}\partial_{s}^{2}h_{kl}^{q}\xi^{k}\xi^{l} =\lambda
(2\pi)^{-\frac{3}{2}}\int d{\bf
k}dsh_{\alpha}^{q}f^{\alpha},\end{array}
\end{displaymath}
where $f^{\alpha}=e_{rl}^{\alpha}f^{rl}$.

 We consider a solution of the Schr\"odinger equation (28) ($V=0$) in
the Gaussian form (if we did not split the gravitational field
into $h^{w}$ and $h^{q}$  then the classical part would be
obtained from the term $Jh$ in eq.(54))
\begin{equation}
\psi^{g}_{t}(h)=\exp\Big(\frac{i}{2\hbar}\int d{\bf
k}h^{q}_{\alpha}\Gamma(t)^{\alpha} h^{q}_{\alpha}\Big).
\end{equation}
In eq.(100)  $(\Gamma^{\alpha}-\Gamma^{*\alpha})^{-1}$ has a
physical meaning as the squeezing of the amplitude $h^{\alpha}$ in
the uncertainty relations (squeezed states are produced during
inflation\cite{aa}\cite{ks}). In general, $\Gamma^{+}$ and
$\Gamma^{\times}$
 are independent solutions of eq.(57). We restrict our discussion
 to the case $\Gamma^{+}=\Gamma^{\times}\equiv \Gamma$ ( and $u_{+}=u_{\times}$).
 Then, the gravitational field resulting from the state (100) is covariant under rotations.
 In the expectation
values (80) and (83) we shall have sums of the form
\begin{equation}
f_{s}^{\alpha}f_{s^{\prime}}^{\alpha}=\Lambda_{mn;rl}f^{mn}_{s}f^{rl}_{s^{\prime}}
\end{equation}
which after averaging over ${\bf k}k^{-1}$ will be expressed by
$q^{rl}$ (94) in the way analogous to the thermal case of sec.7
($\Lambda_{mn;rl}f^{mn}_{s}f^{rl}_{s^{\prime}}\rightarrow
<\Lambda_{mn;rl}>f^{mn}_{s}f^{rl}_{s^{\prime}}$). In detail
\begin{equation}\begin{array}{l} \rho_{t}\simeq
\exp\Big(-\frac{i\lambda^{2}m_{0}^{2}}{2\hbar}\int d{\bf
k}\cr\times(\Gamma(t)-\Gamma(t)^{*})^{-1}\Big(\int_{0}^{t}(u_{t}^{-1}u_{s}f^{\alpha}_{s}
 -u_{t}^{*-1}u_{s}^{*}f_{s}^{\prime\alpha})ds\Big)^{2}\cr
 -\frac{i\lambda^{2}m_{0}^{2}}{2\hbar}\frac{4\pi}{5}\int dk \int_{0}^{t}\Big(u_{s}u_{s^{\prime}}(\sigma^{2}+\delta^{2})^{-1}
\cr(\Gamma(t)-\Gamma(max(s,s^{\prime})))q^{rl}_{s}q^{rl}_{s^{\prime}}
\cr-u^{*}_{s}u^{*}_{s^{\prime}}(\sigma^{* 2}+\delta^{* 2})^{-1}
(\Gamma^{*}(t)-\Gamma^{*}(max(s,s^{\prime})))q^{\prime
rl}_{s}q^{\prime
rl}_{s^{\prime}}\Big)dsds^{\prime}\Big).\end{array}
\end{equation}
In eq.(102) the $f_{\alpha}f_{\alpha}$ term of eq.(83) has been
expressed by $q^{rl}$  and  $u(k)$ is defined in eq.(60) with
$\omega=k$ ( the coefficients $\delta$ and $\sigma$ may depend on
$k$).

 For the final result the integral over
$k$ is crucial. Its exact value depends on the complex functions
$\sigma(k)$ and $\delta(k)$ in eq.(60). In the thermal case the
effective action at high temperature in the exponential (93)was
local in time (for a small $\beta$ ) owing to the $k^{-1}$ factor
coming from $\coth(\frac{1}{2}\hbar\beta k)$. We do not have such
a factor here. Nevertheless, we can see that the non-local final
noise can be large owing to the squeezing factor
$(\Gamma-\Gamma^{*})^{-1}$ in eq.(102).Explicitly, this term is
\begin{equation}\begin{array}{l}

\exp\Big(-\frac{i\lambda^{2}m_{0}^{2}}{2\hbar}\frac{8\pi}{5}\int
dk k^{2}\cr\times(\Gamma(t)-\Gamma(t)^{*})^{-1}\int_{0}^{t}dsds^
{\prime}\Big(u_{t}^{-2}u_{s}u_{s^{\prime}}q^{rm}_{s}q^{
rm}_{s^{\prime}}\cr
+u_{t}^{*-2}u_{s}^{*}u_{s^{\prime}}^{*}q^{\prime rm}_{s}q^{\prime
rm}_{s^{\prime}} -
u_{t}^{-1}u_{t}^{*-1}u_{s}^{*}u_{s^{\prime}}q^{\prime
rm}_{s}q^{rm}_{s^{\prime}}\cr
-u_{t}^{-1}u_{t}^{*-1}u_{s}u_{s^{\prime}}^{*}q^{rm}_{s}q^{\prime
rm}_{s^{\prime}}\Big)\Big).
\end{array}
\end{equation}
The exponential in eq.(102) is of the form
\begin{equation}
\exp(
i\alpha_{1}qq+i\alpha_{2}q^{\prime}q^{\prime}+i\alpha_{3}qq^{\prime}).
\end{equation}
When $\frac{\delta}{\sigma}=i$, $\Gamma=ik$ , $u_{s}=\exp(iks)$
then
 in
 the integral (102) we obtain (this is an
 infinite mode version of eq.(80))
\begin{equation}\begin{array}{l}\rho_{t}\simeq
\exp\Big(-\frac{\lambda^{2}m_{0}^{2}}{2\hbar}\frac{4\pi}{5}\int
dkk\int_{0}^{t} dsds^{\prime}\cr \Big(\exp(-ik\vert
s-s^{\prime}\vert)q_{rl}(s)q_{rl}(s^{\prime})\cr +\exp(ik\vert
s-s^{\prime}\vert)
q^{\prime}_{rl}(s)q^{\prime}_{rl}(s^{\prime})+\cr
-\exp(-ik(s+s^{\prime}))(q_{rl}(s))q^{\prime}_{rl}(s^{\prime})+q^{\prime}_{rl}(s)q_{rl}(s^{\prime})\Big)\Big).
\end{array}
\end{equation}
Representing $\sin(kt)$ as $-k^{-1}\partial_{t}\cos(kt)$  we
integrate over $k$ obtaining  $\partial_{s}\delta(s-s^{\prime})$
in a similar
  way as we did it in the thermal case arriving at
the phase factor \begin{equation}
\exp\Big(-i\frac{\gamma}{2\hbar}\int_{0}^{t}(q_{rl}-q_{rl}^{\prime})
\partial_{s}(q_{rl}(s)+q_{rl}^{\prime}(s)\Big).
\end{equation}
 We obtain the same lhs of the
stochastic equation (98) as in the thermal state , but the noise
resulting from eq.(105)is different than the one of eq.(98) (
non-local in time and non-Markovian).

In eq.(102) the exponential of  the density matrix can be written
in the form
\begin{equation}\begin{array}{l}\rho_{t}\simeq
\int D\xi
D\xi^{\prime}\exp\Big(\frac{i}{2\hbar}\int_{0}^{t}ds(m_{0}\frac{d\xi_{r}}{ds}\frac{d\xi_{r}}{ds}
-m_{0}\frac{d\xi_{r}^{\prime}}{ds}\frac{d\xi_{r}^{\prime}}{ds}
\cr+2V(\xi)-2V(\xi^{\prime})) +i\int
dsds^{\prime}\Big((q^{rl}-q^{\prime rl})C(q^{rl}+q^{\prime
rl})\cr+(q^{rl}+q^{\prime rl})A(q^{rl}+q^{\prime
rl})+(q^{rl}-q^{\prime rl})B(q^{rl}-q^{\prime rl})\Big).
\end{array}\end{equation}
If the exponential (102)-(103) is written in the form (104) then
$\alpha_{1}=A+B+C$, $\alpha_{2}=A+B-C$, $\alpha_{3}=2(A-B-C)$. The
functions $A,B, C$ can be read from eqs.(102)-(103).
 Expanding in $y$
we obtain the non-Markovian stochastic equation as derived in
eq.(84). In eq.(107) the $C$-term is proportional to $y$ whereas
the $A$ and $B$ terms are quadratic in $y$. So the $C$ term gives
the modification of the equation for the geodesic deviation
whereas the $A,B$ terms contribute to the noise.

Eq.(102) simplifies if $\Gamma(t)\simeq const$. Let us set in
eq.(60) $a\delta=i\sigma $ ( $a$ may depend on $k$). Then
$\Gamma(0)=ika^{-1}$. We have a real Gaussian function in eq.(100)
as an initial state. This squeezed state can  still be
approximated by an initial  real function if $a$ is large and
$(kt)^{-1}>>a>>kt$ with $kt<<1$ . In such a case  to eq.(102) only
the term (103) contributes, where $u_{s}\simeq \sigma\cos(ks)$.
Hence
\begin{equation}\begin{array}{l} \rho_{t}\simeq\cr
\exp\Big(-\frac{\lambda^{2}m_{0}^{2}}{4\hbar}\int d{\bf
k}\frac{a}{k}\Big(\int_{0}^{t}(\cos(kt))^{-1}\cos(ks)(f^{\alpha}_{s}
 -f_{s}^{\prime\alpha})ds\Big)^{2}\Big).\end{array}
\end{equation}
The integration over the angles  $k^{-1}{\bf k}$ of
$\epsilon_{rl}^{\alpha}\epsilon_{mn}^{\alpha}$ is expressed by
$<\Lambda_{rl;mn}>$ (92). Hence, finally

\begin{equation}\begin{array}{l} \rho_{t}\simeq
\exp\Big(-\frac{\lambda^{2}m_{0}^{2}}{16\hbar}\int dkk
a\Big(\int_{0}^{t}dsds^{\prime}<\Lambda_{rl;mn}>\cr(\cos(kt))^{-2}\cos(ks)\cos(ks^{\prime})
 \cr\frac{d^{2}}{ds^{2}}(X^{r}y^{l}+X^{l}y^{r})\frac{d^{2}}{ds^{\prime 2}}(X^{m}y^{n}+X^{n}y^{m})\Big).\end{array}
\end{equation}
Eq.(109) gives the spectrum of the noise of the quadrupole
$q^{rl}$ as $8\pi G a k$ ( defined in \cite{wilczek3}) or the
spectrum of the noise of the measured coordinate $\xi$ (if we
perform the differentiation by parts over $ s$ contained in
$q^{rl}$ ) as $8\pi G a k^{5}$ in agreement with
\cite{soda}(eqs.(3.13)-(3.14)). The kinematic form of the noise
(109) is the same as the one for the thermal noise discusssed at
the end of sec.7 and in the Appendix. However, because of a
different spectrum of the noise (109) it is not of the local form
(99).

\section{Summary}We have calculated  the density
matrix resulting from an average over the gravitational field in a
thermal state and in a general Gaussian state. The result shows
that for gravitons in high temperature or in a highly squeezed
state the modification of the geodesic deviation equation applied
in a gravitational wave detection can have large noise amplitude.
In general, the perturbation of the geodesic deviation equation is
non-local and non-Markovian. Our results on the stochastic
geodesic deviation equation show some differences in comparison to
PWZ\cite{wilczek} which may come from
 approximations used by those authors. The formula
for the noise (although quite complicated) can be useful in order
to distinguish the contribution of the graviton noise from other
sources of noise in the gravitational wave detection. Using
eq.(69) we can calculate the transition probability between the
initial and final states of the detector. In the expansion in
$\hbar$ the calculation is reduced to  expectation values of the
noise. The noise could be detectable if gravitational waves come
from an inflationary stage of universe evolution (squeezing) or
from a merge of hot neutron stars.

{\bf Remark} After a submission to arXiv of the first version of
this paper there  appeared extended versions
\cite{wilczek2}\cite{wilczek3} of PWZ paper \cite{wilczek}. Their
modification  of the deviation equation (Eq.(125) of ref.\cite{wilczek3} ) is different from
(98). The difference comes from the approximations discussed in
\cite{wilczek3} at the end of sec.2 where it is assumed that only
one quantum mode of $\xi^{r}$ is excited.

{\bf Acknowledgement} The author thanks an anonymous referee for
some useful suggestions for improvements of the first version of
this paper.
\section{Appendix:High temperature (classical) limit}
We can treat the system in an environment of thermal gravitons by
means of the same procedure as applied  in\cite{habaepj}. The
equation for the gravitational field resulting from the Lagrangian
(85) is
\begin{equation}
\frac{d^{2}h^{rl}({\bf k})}{dt^{2}}+k^{2}h^{rl}({\bf
k})=(2\pi)^{-\frac{3}{2}}\lambda f^{rl}.
\end{equation}
 We write the  solution of eq.(110) (assuming that when $t\leq t_{0}$, where $t_{0}$ is the initial time,  $h_{rl}$
 behaves as free wave) in the form
(in the transverse-traceless
gauge)\begin{equation}\begin{array}{l} h_{rl}({\bf
k})=h_{rl}^{w}({\bf k})+h_{rl}^{th}({\bf k})+h_{rl}^{I}({\bf
k})\cr=h_{rl}^{w}+e_{rl}^{\alpha}(h_{0}^{\alpha}\cos(kt)+k^{-1}\Pi_{0}^{\alpha}\sin(kt))
\cr+\lambda (2\pi)^{-\frac{3}{2}}
\Lambda_{rl;mn}\int_{t_{0}}^{t}k^{-1}\sin(k(t-t^{\prime}))f_{mn}(t^{\prime})dt^{\prime},
\end{array}\end{equation}
here
\begin{displaymath}
h^{w}_{rl}(t,{\bf x}) =(2\pi)^{-\frac{3}{2}}\int d{\bf
k}\exp(i{\bf kx})h^{w}_{rl}({\bf k},t)
\end{displaymath}
is the gravitational wave, $h^{th}$ describes the thermal modes
distributed with the Gibbs equilibrium measure, $h_{0}^{\alpha}$
and $\Pi_{0}^{\alpha}$ are random initial conditions. $h_{rl}^{I}$
is the gravitational field created by the motion of $m_{0}$.
$h_{rl}^{w}$ and $h_{rl}^{th}$ satisfy the homogeneous equation
\begin{displaymath}
\frac{d^{2}h_{rl}({\bf k})}{dt^{2}}+k^{2}h_{rl}({\bf k})=0.
\end{displaymath} The equation of motion for the coordinate $\xi$ is

\begin{equation}\begin{array}{l}
\frac{d^{2}\xi_{r}}{dt^{2}}=\frac{\lambda}{2}\frac{d^{2}h^{w}_{rl}({\bf
        k})}{dt^{2}}\xi^{l}
+\frac{\lambda}{2}(2\pi)^{-\frac{3}{2}}\int d{\bf
k}\frac{d^{2}h_{rl}^{th}({\bf k})}{dt^{2}}\xi^{l}\cr
+\frac{\lambda}{2}(2\pi)^{-\frac{3}{2}}\int d{\bf
k}\frac{d^{2}h_{rl}^{I}({\bf k})}{dt^{2}}\xi^{l}.
\end{array}\end{equation}

 We insert the gravitational field from eq.(111) into eq.(112)
. We obtain
\begin{equation}
\frac{d^{2}\xi_{r}}{dt^{2}}=\frac{\lambda}{2}\frac{d^{2}h^{w}_{rl}({\bf
        k})}{dt^{2}}\xi^{l}+N_{rl}(t)\xi^{l}+F_{r}(\xi,t),
\end{equation}
where F is a non-linear interaction resulting from the interaction
with the environment and\begin{equation}\begin{array}{l}
N^{rl}(t)=\int d{\bf k}
N^{rl}(k,t)=-\frac{\lambda}{2}(2\pi)^{-\frac{3}{2}}\int d{\bf k}
k^{2}e^{\alpha}_{rl}\cr
(h_{0}^{\alpha}\cos(kt)+k^{-1}\Pi_{0}^{\alpha}\sin(kt)).
\end{array}\end{equation} The classical Gibbs distribution (a classical
limit of the quantum Gibbs distribution of sec.7) is
($\Pi_{\alpha}$ has the meaning of the initial canonical momentum)
\begin{equation}
d\Pi_{0}^{\alpha}dh_{0}^{\alpha}\exp\Big(-\frac{\beta}{2}\int
d{\bf k}(\vert\Pi_{\alpha}\vert^{2}+k^{2}\vert
h_{\alpha}\vert^{2})\Big).
\end{equation}
Calculating the correlation functions of the noise (assuming the
Gibbs distribution of the initial values) we obtain
\begin{equation}\begin{array}{l}
\large< N^{rl}({\bf k},t) N^{mn}({\bf
k}^{\prime},t^{\prime})\large>=\frac{\lambda^{2}}{4}\beta^{-1}(2\pi)^{-3}\delta({\bf
k}+{\bf k}^{\prime})k^{2}\Lambda_{rl:mn}\cr\cos(k(t-t^{\prime})).
\end{array}\end{equation}
So that \begin{equation}\large< N^{rl}(t)
N^{mn}(t^{\prime})\large>=<\Lambda_{rl:mn}>\frac{\lambda^{2}}{4\pi}
\beta^{-1}\partial_{t}^{2}\partial_{t^{\prime}}^{2}\delta(t-t^{\prime}).
\end{equation}
The non-linear force resulting from the graviton environment is
\begin{equation}\begin{array}{l}
F^{r}=\frac{1}{2}\lambda^{2}(2\pi)^{-3}\partial_{t}^{2}\Big(\int
d{\bf k}\cr
\Lambda_{rl;mn}\int_{t_{0}}^{t}k^{-1}\sin(k(t-t^{\prime}))f_{mn}(t^{\prime})dt^{\prime}\Big)\xi^{l}
\cr =\frac{1}{2}\lambda^{2}m_{0}(2\pi)^{-3}\int dk k^{2}\cr
<\Lambda_{rl;mn}>(\int_{t_{0}}^{t}\partial_{t}\cos(k(t-t^{\prime}))f_{mn}(t^{\prime})dt^{\prime}\xi^{l}
+f_{mn}(t)\xi^{l}).
\end{array}\end{equation}
We write the last factor as\begin{equation}\begin{array}{l}
(\int_{t_{0}}^{t}\partial_{t}\cos(k(t-t^{\prime}))f_{mn}(t^{\prime})dt^{\prime}\xi^{l}(t)
+f_{mn}(t)\xi^{l}(t))\cr=(-\int_{t_{0}}^{t}\partial_{t^{\prime}}\cos(k(t-t^{\prime}))f_{mn}(t^{\prime})dt^{\prime}\xi^{l}(t)
+f^{mn}(t)\xi^{l}(t))\cr=\int_{t_{0}}^{t}\cos(k(t-t^{\prime}))\partial_{t^{\prime}}f_{mn}(t^{\prime})dt^{\prime}\xi^{l}(t)\cr+\cos(k(t-t_{0}))f_{mn}(t_{0})\xi^{l}(t).
\end{array}\end{equation}
Then
\begin{equation}\begin{array}{l}
\int_{t_{0}}^{t}\int dk
k^{2}\cos(k(t-t^{\prime}))\partial_{t^{\prime}}f_{mn}(t^{\prime})dt^{\prime}\cr=-2\pi\int_{t_{0}}^{t}\partial_{t^{\prime}}^{2}
\delta(t-t^{\prime})\partial_{t^{\prime}}f_{mn}(t^{\prime})dt^{\prime}
=2\pi\partial_{t}^{3}f_{mn}(t),\end{array}
\end{equation}
where we assumed that  at $t_{0}$ the first and the second
derivatives of  $f_{mn}$  are zero. If we assume that in (119)
$f_{mn}(t_{0})=0$ then  we can write eq.(113) in the form
\begin{equation}\begin{array}{l}
\frac{d^{2}\xi^{r}}{dt^{2}}=\frac{\lambda}{2}
\frac{d^{2}h_{w}^{rl}}{dt^{2}}\xi_{l}\cr-\frac{\lambda^{2}}{5\pi}(\delta_{rm}\delta_{ln}-\frac{1}{3}\delta_{rl}\delta_{mn})
\xi^{l}\partial_{t}^{3}f^{mn} +N^{rl}(t)\xi_{l}.
\end{array}\end{equation}
Eq.(121) coincides with eq.(98) as the noise (117) is the same as
the one defined by eqs.(98)-(99).

\end{document}